# Modelling cellular signalling variability based on single-cell data: the TGFβ-SMAD signaling pathway

Uddipan Sarma[2,3*], Lorenz Hexemer[2,*], Uchenna Alex Anyaegbunam[2,*], Stefan Legewie[1,2,#]

**Affiliations:**
[1] University of Stuttgart, Department of Systems Biology, Allmandring 30E, 70569 Stuttgart, Germany
[2] Institute of Molecular Biology (IMB), Ackermannweg 4, 55128 Mainz, Germany
[3] Vantage Research, Sivasamy St, CIT Colony, Mylapore, Chennai, Tamil Nadu 600004. India.

[*] contributed equally
[#] correspondence: s.legewie@imb-mainz.de

## ABSTRACT

Non-genetic heterogeneity is key to cellular decisions, as even genetically identical cells respond in very different ways to the same external stimulus, e.g., during cell differentiation or therapeutic treatment of disease. Strong heterogeneity is typically already observed at the level of signaling pathways that are the first sensors of external inputs and transmit information to the nucleus where decisions are made. Since heterogeneity arises from random fluctuations of cellular components, mathematical models are required to fully describe the phenomenon and to understand the dynamics of heterogeneous cell populations. Here, we review the experimental and theoretical literature on cellular signaling heterogeneity, with special focus on the TGFβ/SMAD signaling pathway.

**Running title:** Modeling cellular signaling variability

# INTRODUCTION

Heterogeneity is an implicit part of life that is observed on nearly all biological scales [1]. Perhaps the most widespread effect of heterogeneity is the creation of a wide spectrum of lifeforms during evolution [2]. Primarily, heterogeneity aids robustness to a biological system in a fluctuating environment, such that, a broader niche in phenotypic traits is collectively achieved by a population of a species, which may then add to better chances of their survival [3]. However, the presence of heterogeneity is observed at even finer levels - individuals within one species exhibit unique properties, and further, genetically identical cells within the same (multicellular) organism utilizes heterogeneity to achieve distinct cell fate decisions [1]. At the level of single cells heterogeneity in expression of individual gene or protein (also called cell-to-cell variability) may lead to cell-specific responses to external cues and distinct physiological trajectories [1, 3].

In this chapter, we provide an overview over cellular heterogeneity in the context of intracellular signaling. We will specifically discuss how a signaling in a heterogeneous cell population can be modelled *in silico* and will point out assumptions underlying these models. The chapter is divided in three parts: In the first Section, we describe biological systems where signaling heterogeneity plays a role in cellular decision making. Furthermore, we review evidence that signaling heterogeneity is mostly deterministic in nature and discuss the molecular sources of signaling heterogeneity. In the second part, we describe mathematical modeling frameworks that can be used to model heterogeneous cell populations and fluctuations in cellular signaling pathways. We mainly focus on deterministic modeling approaches using ordinary differential equations, in which heterogeneity is introduced by parameter sampling and discuss approaches for the quantitative fitting of single-cell models to experimental data. In the third part, we focus on the TGF/SMAD signaling pathway that plays a key role in tissue homoeostasis and cell migration, but also in diseases such as cancer. We review the literature on single-cell analysis of this pathway, and demonstrate that key features of heterogeneous TGFβ/SMAD signaling can be understood by mechanistic modeling. We then discuss our recent modeling work, in which we quantitatively described cellular subpopulations of TGFβ/SMAD signaling and heterogenous signaling at the single-cell level. As an outlook for future research, we summarize how fluctuations in signaling pathways affect noisy downstream gene expression and decision making.

## 1. Relevance and features of cellular signaling heterogeneity

### 1.1. Heterogeneity in cellular responses to external cues

From early microscopic observations in cell culture, it became clear that not all cells respond identically to the same external stimulus. Intriguingly, it seems that not only genetic differences between cells contribute to heterogeneity, but that non-genetic origins arising from stochasticity in cellular networks also play an important role. In recent years, evidence has accumulated demonstrating that genetically identical cells show differences in differentiation programs [4], drug resistance [5, 6], and viral pathogenesis [7]. In the following paragraphs, we provide a brief summary of biological systems where heterogeneity plays a functional role in cellular behavior.

***Stress responses in unicellular organisms:*** Heterogeneity is an important part of cellular decision making, as evidenced by stochastic cellular differentiation events, where parts of a cell population randomly enter a new fate. Well-known examples include bacterial stress responses (e.g., [8]). For instance, during bacterial competence, external stress induces a regulatory program in *Bacillus subtilis* that allows cells to take up DNA from the environment, thereby priming them for adaptation to stress conditions. In line with a stochastic event, it was shown that the decision to become competent is dictated by random fluctuations in the master transcriptional regulator ComK [8]. In some cases, stochastic fluctuations in stress networks occur constitutively, i.e., even in the absence of external stress. This phenomenon, known as bet hedging, ensures that subsets of cell populations are prepared to rapidly respond to stressors, thereby ensuring survival of the population in case of external changes. In *Saccharomyces cerevisiae,* isogenic clonal populations display a range of growth rates and slow growth predicts resistance to heat killing. At the molecular level, Tsl1, a trehalose-synthesis regulator, is a key component of the observed resistance, and cell-to-cell variability in Tsl1 expression correlates with growth rate and predicts cellular survival in response to stress [9].

***Cellular differentiation:*** Likewise, in eukaryotes, cell fate decisions during tissue development appear to occur stochastically in genetically identical cell populations, and this is thought to allow for a disversificvation of tissues. At the molecular level, evidence is accumulating that random fluctuations in the levels of signaling pathways and master transcription factors govern cell fates (Fig. 1A). For instance, in *Drosophila melanogaster* cell-to-cell variability in the expression of the transcription factor spineless creates the retinal mosaic for colour vision and is thus important for the spatio-temporal organization of the eye [10]. Moreover, subpopulations of clonally derived hematopoietic progenitor cells with low or high expression of a stem cell marker (Sca-1) were observed to be in dramatically different transcriptional states and gave rise to different blood cell lineages in multipotent murine hematopoietic cell line [4]. In developing mice, cell-to-cell variability in the expression of certain genes (e.g., Fgf4) was found to determine the inner cell mass (ICM) lineage segregation of the blastocyst [11]. Further evidence for non-genetic heterogeneity and its impact on animal development comes from genetic mutations with incomplete penetrance. These mutations cause physiological defects only in a subset of genetically identical animals. For instance, during *C. elegans* development, the expression level of elt-2, a self-activating transcription factor is critical for intestinal cell-fate specification. Strong embryo-to-embryo heterogeneity in elt-2 expression and thus failure of intestinal development in a subset of embryos is observed if the upstream regulator skn-1 is inactivated by a mutation. Thus, the skn-1 mutant shows incomplete penetrance, likely because the lowered input signal shifts heterogeneous elt-2 expression to a range, where its fluctuations have profound effects on intestinal development [12] .

***Tumor progression and drug resistance***: Cancer cell therapy aims for a complete eradication of tumor cells. However, in reality, complete killing is rarely achieved, as signaling pathways involved in the execution of cell death fail to be activated in all cells and certain subpopulations are therefore resistant to the therapeutic treatment (Fig. 1B). Thus, cytotoxic drugs often result in fractional killing, especially when the drug concentration inside a tumor is limiting. In cell culture, fractional killing has been reported in response to a variety of treatments, including chemotherapy and apoptosis-inducing receptor ligands such as TRAIL [5, 6, 13–16]. At the molecular level, fractional killing involves cell-to-cell variation in cellular regulatory molecules, e.g., in the tumor suppressor p53 in response to chemotherapy [6][. Thus, non-genetic variability may confer resistance to therapeutic intervention and could play a role in tumor evolution [17].

These examples demonstrate that cellular heterogeneity can have a strong impact of cell fate decisions in biological systems. It is therefore crucial to quantitatively measure cellular heterogeneity using experimental methods such as live-cell imaging, flow cytometry or single-cell RNA sequencing (reviewed in [18–23]). Furthermore, predictive modeling approaches describing the variability of molecular networks at the single-cell level will be essential to rationally manipulate cellular differentiation processes and to design effective combinatorial therapeutic intervention strategies.

## 1.2. Heterogeneity in the activity of cellular signaling pathways

The above examples of stochastic decision making mainly focused on cellular fluctuations in nuclear transcription factors. Signaling pathways controlling these master transcription factors similarly show strong non-genetic cell-to-cell variation that is linked to cell fate. This was initially shown for mitogen-activated protein kinase (MAPK) signaling and later extended to other signaling systems.

In a pioneering study, the group of James Ferrell analyzed Xenopus oocyte maturation in response to the maturation-inducing hormone progesterone [24]. This maturation response is mediated by the MAPK signaling pathway and due to the large size oocytes the authors could perform single-cell Western Blot experiments to determine cell-to-cell variation in the activity of this pathway. They showed that MAPK signaling was activated in an all-or-none manner within individual cells, i.e., every oocyte either had low or high (but not intermediate) MAPK activity level, and MAPK activity therefore showed a bimodal distribution (Fig. 1C, right). This switch-like (ON or OFF) MAPK activation with strong heterogeneity between cells caused maturation in only a fraction of cells, and ON/OFF-switching was shown to arise from a positive feedback at the level of the signaling pathway [24]. Likewise, switch-like and heterogeneous MAPK activation was reported to be involved in other cell fate decisions such as T cell activation [25], PC12 cell differentiation [26] and in the UV stress response mediated by the closely related JNK MAPK signaling pathway [27]. Thus, MAPK signaling frequently mediates switch-like and highly heterogeneous cell fate decisions. However, the pathway can also be gradually activated, with a unimodal but heterogeneous distribution of activity states in the population (Fig. 1C, left), e.g., in EGF-stimulated fibroblasts. There, the pathway transmits quantitative information to the nucleus, and switch-like cell fate decisions are established at the level of downstream gene-regulatory networks [28, 29] In conclusion, heterogeneity in MAPK signaling is a widespread phenomenon, but the qualitative features (uni- vs. bimodal distribution) of the heterogeneous population can vary, implying plasticity in network behavior.

Other signaling systems seem to be less flexible in their signaling output. For instance, the apoptosis signaling system, involved in sensing cytotoxic stress and death receptor signals, seems to invariably induce heterogeneous all-or-none responses at the level of the executing caspase enzymes (e.g., [30]). This may ensure that programmed cell death is executed completely and irreversibly, but only a fraction of cells in a tissue, thereby preventing complete loss of all cells in a tissue. Yet other signaling pathways such as Akt [31, 32], and TGFβ/SMAD [33, 34] signaling show a heterogeneous, but gradual (unimodal) response in the majority of cellular systems studied. Thus, these signaling pathways transmit quantitative information about extracellular stimulus concentration from the cell membrane to the nucleus, where gene-regulatory networks mediate cell fate decisions.

One important question in signal transduction is how accurate information transmission is possible despite strong heterogeneity in the activity of gradual signaling pathways. Based on a combination of quantitative experiments, mathematical modeling and concepts from information theory it became clear that quantitative information is frequently encoded in the temporal dynamics (i.e., the shape) of the signal [35–38]. For instance, the NFkB signaling pathway shows oscillatory dynamics and the nature of activated target genes in the nucleus depends on the frequency and amplitude of the signaling pathway oscillation [39]. Another concept for reliable signal transmision despite signaling heterogeneity is relative signal transmission, where the (noisy) absolute signaling activity is less important for cellular behavior when compared to the (more robust) stimulus-induced fold-change over basal [29, 40–43]. Combined live-cell imaging and modeling studies support this concept for Erk, Wnt and TGFβ signaling pathways, and it has also been described how downstream gene-regulatory networks in the nucleus could respond to fold-changes rather than absolute signaling levels [40–42].

Taken together, both gradual and bimodal signaling pathways show strong cell-to-cell variability in their activity. Single-cell experiments characterize quantitative information transmission and switch-like decision making in a heterogeneous cell population and therefore provide a basis for quantitative modeling.

### 1.3 Signaling fluctuations are often non-genetic and temporally stable

To model signaling heterogeneity, assumptions need to be made about the properties and origins of the fluctuations. Evidence from the literature suggests that signaling heterogeneity is non-genetic and that the cell-specific features of signaling can be assumed to be temporally stable during a typical cellular stimulation experiment.

Cell cultures are often derived from tumor cells and are thus potentially genetically unstable. This raises the question of whether heterogeneity in signaling pathways is really non-genetic in nature. Strong evidence for a non-genetic contribution to heterogeneity in signaling events comes from re-stimulation experiments with ligands and drugs triggering cell death (Fig. 1B). In several of these studies, an initial treatment killed the majority of the cell population and the cells were kept in culture for several days before they were subjected to another treatment with the same stimulus. Even though the initial survivors were resistant to the first treatment a large part of them became sensitive to the second treatment (Fig. 1B). Thus, resistance was not inherited to all offspring of resistant cells, i.e., genetically determined, but was gradually lost during several days of culture, arguing for an epigenetic mechanism of heterogeneity [5, 13, 44, 45].

A second line of evidence for a non-genetic signaling heterogeneity came from sister cell experiments, in which cells and their division events were tracked before stimulation to record cellular progeny (Fig. 1D). A common observation in these lineage tracing experiments is that freshly divided sister cells show very similar signaling responses, and that the similarity of sisters got lost over time after the common division event. For instance, for TRAIL-induced apoptosis, it was shown that freshly divided sister cells have a high correlation in the death time, i.e. the time it takes for the TRAIL stimulus to induce cell death [46]. Thus, the experiment suggests that sister cells are initially in a very similar (signaling) state, but interestingly they loose the signaling similarity with a characteristic half-life of 11h (after the common division event). Similar observations were made for other regulatory networks, including the cell cycle, the spindle assembly checkpoint, MAPK as well TGFβ signaling [33, 47–50]. Sister cell experiments have several important implications for cellular signaling heterogeneity, and mathematical modeling thereof:

1. First, they further support non-genetic sources of heterogeneity, since the loss of sister cells similarity on a time scale of hours is much faster than any genetic drift due to DNA mutations.

2. Second, the initially high sister cell similarity suggests that heterogeneity does not arise from stochastic dynamics in signaling reactions: If signaling molecules are present in very low amounts, signaling reactions (e.g., phosphorylation) could be probabilistic events that give rise to strong heterogeneity and high dissimilarity in the signaling events of freshly divided sisters (or even high dissimilarity if the same cell would be stimulated repeatedly). Certain signaling pathways indeed show such stochastic dynamics (see Section 2.3 and [51]), but the sister cell experiments suggest that this is typically not the case, which agrees with the fact that signaling proteins are often expressed at high molecule numbers [52].

3. Third, the time scale of sister cell similarity (multiple hours) suggests that signaling heterogeneity can be assumed to be stable at the time scale of a typical stimulation experiment (minutes to hours). The conclusion of temporal stability is also supported by a recent restimulation analysis with insulin-like growth factor, in which the rank of single-cells with respect to Akt signaling activity was stable over several hours [32] .

These conclusions greatly simplify the mathematical modeling of cellular signaling pathways, as stochastic modeling of signaling dynamics can be neglected in most cases. Instead, a deterministic ordinary differential equation approach can be used, in which certain kinetic parameters are assumed to be different between cells, but stable over time.

### 1.4. Signaling heterogeneity arises from fluctuations in signaling protein levels

To introduce heterogeneity into a mathematical model, we need to know the molecular sources of temporally stable signaling pathway fluctuations. Obvious sources for signaling fluctuations are cell-to-cell differences in cell cycle stage [19] or cell density [53]. In growing adherent mammalian cells, cell divisions combined with cell motility can create variations in local cell densities, cell–cell contacts, relative location, and the amount of free space per cell. Combined, these parameters constitute the population context of an individual cell [54, 55]. However, single-cell signaling studies exist where these sources have been excluded experimentally, but the heterogeneity in the signaling output persists [33, 48, 51].

In recent years, it has become apparent that random fluctuations in the total cellular concentrations of signaling proteins may underlie temporally stable cell-to-cell variability in signaling pathway activity (Fig. 1E). In some cases, signaling heterogeneity could be traced back to fluctuations in the expression of specific signaling molecules, e.g., in MAPK [25], PI3K/Akt [56] and JAK/STAT [57] signaling. For instance, a pioneering work on T cell activation showed that MAPK activation depends on cell-to-cell variability in the expression of the CD8 co-receptor and the antagonizing phosphatase SHP-1. Interestingly, even though each protein contributes to variability, co-regulation of CD8 and SHP-1 expression levels at the single-cell level limits diversity and promotes robustness of signaling [25]. In other systems, signaling heterogeneity cannot be explained by fluctuations in a single protein, but the control seems distributed over many protein levels. For example, in their work on apoptosis signaling, Spencer et al showed that no single protein level could accurately predict cell death timing in response to TRAIL treatment, unless a specific pathway regulator (BID) was overexpressed and its cellular level then became a good predictor of cell death timing [46].

Even in genetically identical cell populations, (signaling) proteins show strong cell-to-cell variability in their levels, since epigenetic control and gene expression are stochastic events at the single-cell level (see also Section 2.3): Each gene is present only at two copies (alleles) per cell and gene regulation at such low molecule numbers is probabilistic [58–61]. As a consequence, each (signaling) protein level follows a log-normal distribution, that is, the fold-change of a protein is normally distributed around its population mean [46, 58, 62]. Although the mean and standard deviation of the distribution are protein-specific, a typical human protein shows a three-fold difference in expression across the cells of a population (where the fold-change is measured between the 10$^{th}$ and 90$^{th}$ precentile of the distribution) [47, 58]. This strong variation in each and every signaling protein leads to strong fluctuations in signaling. Time-resolved measurements further suggest that fluctuations in signaling protein expression and pathway activity are coupled: when protein expression fluctuations in mammalian cells are followed over time, the time scale of stochastic changes in protein expression is similar to the time scale of signaling pathway desynchronization between sister cells (multiple hours to days) [46, 58]. This is consistent with a model, in which sister cells initially share the same proteome content and signaling activity, and then over time and simultaneously loose both types of similarity due to stochastic gene expression fluctuations [46, 47, 58].

Taken together, these observations indicate that mammalian signaling proteins show strong fluctuations in their levels. In the following, we will discuss deterministic modeling approaches of cellular signaling pathways in which cell-to-cell variability in signaling protein concentrations is taken into account.

# METHODS

## 2. Mathematical modeling of cellular heterogeneity

Signaling pathways often respond in very similar manner to stimulation when the same cell is stimulated repeatedly, or when sister cells are in a similar state and harbor a similar proteome content. These observations suggest that cells respond deterministically to stimulation and that deterministic mathematical modeling approaches can be used to simulate cellular heterogeneity *in silico*. Such deterministic models are often based on ordinary differential equations (ODEs) which represent reaction networks within the cell, typically using mass action-based reaction kinetics (reviewed in [63]). ODE models assume that the biochemical molecules in the cell are present in sufficiently large amounts (and well-stirred), so that stochastic fluctuations at the single-molecule level can be neglected and the biochemical species can be described using continuous variables, representing average molar signaling protein concentrations within the cell. If an ODE system is simulated twice with the same set of kinetic reaction parameters and initial conditions (i.e., protein concentrations) it will yield exactly the same solution, representing the deterministic nature of the apporach. This determinism is in contrast to stochastic simulation algorithms (reviewed in [64]) which explicitly describe single-molecule fluctuations and therefore give rise to distinct simulation results for each realization.

In this Section, we discuss how intracellular signal transduction in a heterogeneous cell population can be modelled *in silico*. We mainly focus on deterministic ODE-based modeling, and point out how these models can provide insights into several aspects of cellular heterogeneity, including noisy decision making and the robustness of signaling networks. We then discuss how these models can be calibrated based on single-cell data to obtain a quantitative match between experiment and theory. Finally, we briefly summarize stochastic modeling approaches that are relevant for the modeling for some signaling networks operating at low molecule numbers and particularly for gene-regulatory networks involved in cell fate decisions downstream of signaling pathways.

### 2.1 Applications and limitations of population-average models

Experiments aimed at understanding intracellular processes were tradionally performed in bulk, combining material from thousands to millions of cells. For instance, signaling pathway dynamics were studied using Western Blot experiments, in which phosphorylated signaling intermediates are detected using phospho-specific antibodies [24, 65]. Due to averaging over a large number of cells, these experiments do not provide information about single-cell heterogeneity, but only represent the behavior of one hypothetical average cell. In systems biology, early quantitative models were built based on the available population-average data and therefore describe one representative cell (Fig. 2A).

Even though not meant to represent cellular heterogeneity, population-average models can provide important insights into several signaling phenomena at the single-cell level including cellular decision making and robustness of networks against fluctuations in their components. For instance, Ferrell et al (1998) showed that the decision of Xenopus oocyte maturation in response to progesterone involves an all-or-none biological response at the level of MAPK signaling (see Section 1.2; [24]). To better understand this single-cell phenomenon, the authors constructed a population-average model of the signaling network, and concluded that switch-like (bistable) behavior in the MAPK cascade arises from a positive feedback loop that amplifies the signal once MAPK signaling exceeds a certain threshold. Likewise, other population-average modeling studies provided insights into mechanisms of switch-like decision making and therefore have implications for bimodal behavior at the single-cell level [26, 66–71]. Population-average models further provided insights into biological robustness against fluctuations in signaling protein concentrations [25, 72–75]: For example, using single-cell experiments, Kamenz et al (2015) observed that the timing of mitotic events is highly robust and is buffered against variations in the concentrations of mitotic regulatory proteins [75]. A population-average model could explain the observed robustness and predicted conditions where mitotic timing is compromised. Thereby, the population-average model identified critical transitions in the network and experimental validation showed that these transitions led to network failure in a subset of cells due to cell-to-cell variability in the molecular components. In general, for gaining insights into robustness, a so-called sensitivity analysis can be performed, in which the initial conditions and kinetic parameters are systematically perturbed (usually one at a time) to understand their role in the behavior of the network. Given that cellular signaling fluctuations often arise from cell-to-cell variability in signaling protein expression (Section 1.4), sensitivity analyses focusing on the impact of signaling protein concentrations at the population level can provide valuable insights into biological variability and robustness in single cells [25]. However, though potentially useful, population-average models do not directly represent signaling distributions across single cells (as those shown in Fig. 1C), and therefore do not allow for a quantitative comparison of the model simulations to a single-cell experiment. Furthermore, a population-average model constructed based on population-average data may lead to misleading conclusions for cell fate decision networks with ON/OFF-behavior [24, 62]: In binary decision making, single cells show show all-or-none (digital) signaling, but do so heterogeneously for a given stimulus concentration. Hence, the activity distribution is bimodal (Fig. 1C), and the mean response of the population gradually increases with increasing stimulation. Therefore, at the population-average level, digital responses may appear gradual (analogue) when averaged.

Taken together, population-average models fail to quantitatively describe single-cell distributions and can only be as insightful as the experimental data they are based on. Models that are designed solely based on population-average data implicitely assume that the population-average data is a good approximation for the true underlying single-cell behavior, and therefore may lead to wrong conclusions. To overcome, these limitations, models of cellular heterogeneity were developed to quantitatively describe the cell-to-cell variability in signal transduction.

### 2.2 Deterministic models of signaling heterogeneity – implementation and scope

How does a deterministic model of cellular signaling heterogeneity look like? In most cases, heterogeneity is simulated using an ensemble of single-cell models (Fig. 2A). Here, individual cells are described by repeating the simulation using the same deterministic ODE model, each simulation run corresponding to one cell, and cell-to-cell variability is introduced by perturbing the model in each realization.

Based on the arguments presented in Section 1.4, the perturbation leading to cell-to-cell variability is mainly introduced by assuming cell-specific signaling protein concentrations. Additionally, kinetic parameters can be assumed to be cell-specific if the corresponding reaction rates are in turn controlled by (fluctuating) proteins (this may, for example, be true for enzymatic reactions). Given that protein expression levels are log-normally distributed (see Section 1.4), single-cells are modeled by repeated simulations in which all protein concentrations in the model are sampled from independent log-normal distributions (Fig. 2A). In this approach, sometimes termed Monte-Carlo sampling or non-linear mixed effect modeling, the protein fluctuations are typically restricted to a biologically reasonable range. Specifically, the coefficient of variation of the lognormal distribution (CV = std / mean) is chosen between 0.1 and 0.4 [46, 58, 62]. Since proteins from the same pathway may co-regulated at the single-cell level [58], sometimes correlated fluctuations in signaling protein fluctuations are assumed [33]. Notably, sampling only the initial total protein concentrations and leaving the model otherwise unperturbed, assumes that non-genetic sources of signaling heterogeneity are temporally stable during pathway stimulation. Hence, the modeling framework captures the key features of cellular signaling heterogeneity, including deterministic behavior, temporal stability and protein concentrations as a noise source (Section 1).

Compared to a population-average model, such single-cell ensemble modeling approaches reproduce the complete heterogeneous cell population and allow for a quantitative comparison of the model to experimental single-cell data. Specifically, while the population-average model by definition only represents the mean, ensemble models capture higher momentum statistics of the heterogeneous model species such as the standard deviation, their correlations or time course features such as the autocorrelation function (e.g., [46]). Deterministic ensemble modeling approaches have been used in a several studies on signaling, often in combination with experimental analyses at the single-cell level [14, 15, 76–85, 16, 86, 87, 35, 41, 46, 49, 62, 70, 72]. These models provided experimentally testable predictions and led to a better understanding of heterogeneous signal transduction. In particular, the following phenomena were analyzed (Fig. 2C):

***1) Sources of signaling fluctuations:*** Ensemle models were used characterize how fluctuations in individual signaling protein expression levels affect the signaling outcome. Thereby, the most critical signaling protein expression fluctuations could be identified as main sources of cell-to-cell variability in signal transduction [16, 33, 46, 49, 57, 87–89]. This led to a better understanding of molecular mechansisms causing heterogeneous cellular decision making.

***2) Design principles of biological robustness:*** Robust and reliable signal transduction must occur despite strong noise. By adding or removing certain reactions in the models, insights were gained into design principles that mitigate signaling variability [41, 62, 72, 80], thereby promoting robustness, e.g., during embryonic development [72]. Several robustness-promoting network motifs could be identified including negative feedback [72, 80], fold-change detection [41] and correlated expression fluctuations in positive and negative regulators of signaling pathways [80, 86].

***3) Characterization and manupulation of heterogeneous decision making:*** For cellular differentiation and during stress responses, cell-to-cell variability may be beneficial, as not all cells of a heterogeneous population enter a new fate and die in response to stress, respectively (see Section 1.1). Modeling of signaling pathways involved in cellular differentiation and cell death allowed for a quantitative analyses of decision making at the level of signaling, and thus for the emergence of bimodal signaling distributions. The models yielded predictions for the reprogramming of cell fates for novel experimental conditions [83, 84] and allowed for the optimization of therapeutic treatment responses in cell culture [14–16, 82]

***4) Insights into alternative biological mechanisms and network topologies:*** Since certain network motifs affect the characteristics of biological fluctuations (see above), attempts were made to infer (reverse engineer) the wiring of signaling networks based on single-cell data. The idea is that signaling fluctuations contain a fingerprint for the underlying molecular interactions, and that the model topology that best describes the signaling fluctuations is the most probable one. Several studies used a defined model topology and compared a set of relatively minor modifications in the model against single-cell data [85, 90]. In a less biased top-down approach, Sachs et al inferred the topology of signaling networks from single-cell data without prior knowledge using a Bayesian framework [91].

***5. Integration of single-cell and population-average data:*** Ensemble models of cell populations allow for simulations at both the single-cell and population-average levels, and can thus be used to integrate both types of data [33, 57, 85]. Thereby, the models on the one hand exploit highly informative single-cell data which often can be done only at low throughput and for few molecular species (especially for time-resolved live-cell imaging). On the other hand, they take into account population-average information that can more easily collected for multiple experimental conditions and molecular species. Accordingly, the integration of population-average and single-cell data led to a better discrimination of competing model hypotheses when fitting an emsemble model to experimental data [85].

### 2.3 Stochastic modeling of signaling and gene expression heterogeneity

Signal transduction cascades typically control cellular decisions by activating gene expression responses in the nucleus. Expression of target genes (e.g., cell cycle regulators or cell adhesion molecules) then controls the morphological features of a cell such as cell division and migration. In addition, target genes often act as negative feedback regulators that downregulate the signal once gene expression has been activated [52]. Thus, signaling and gene expression responses are intimately connected, and both may need to be taken into account in realistic models of cellular decision making. In this context, it should be pointed out that determinstic models may no longer be suitable for modeling of cellular heterogeneity if gene expression is taken into account, e.g., for modeling transcriptional feedback or nuclear propagation of the signal.

The reason is that gene regulation is an intrinsically stochastic process with strong temporal fluctuations (reviewed in [59], although deterministic sources of heterogeneity sources (i.e., the cellular state) also seem to play a role [54, 60]. Stochastic behavior arises from the fact that transcriptional regulators are typically expressed at very low levels, and that a cell contains only two copies of each gene. As a result, random (Brownian) fluctuations at the level of individual reactions are not averaged out and significantly impact on the activity of a gene, especially at the level of mRNA production. Therefore, stochastic approaches are typically used for modeling heterogeneity of gene expression [59, 64]. In early work, Arkin and colleagues used the Gillespie algorithm to simulate biochemical reactions leading to gene expression, and predicted stochastic cell-to-cell variation in protein numbers for biologically realistic parameter ranges [92, 93]. The prediction of stochastic mRNA and protein expression was later confirmed experimentally in bacteria and mammalian cells (reviewed in [59]). In higher organisms, noise seems to be larger in magnitude compared to bacteria, since chromatin states seem to give rise to switching of genes between ON and OFF states. In time courses of single-cell gene expression, this is observable as transcriptional bursts, i.e., episodes of high gene expression that are separated by phases with low activity [60, 61, 94, 95]. The simplest stochastic model which realistically describes transcriptional bursts is the so-called random telegraph model, where a gene promoter is assumed to reversibly switch between a transcriptional active and an inactive state (reviewed in [59]). Notably, depending on the gene under consideration more promoter states may need to be taken into account to describe the data [94, 96]. To jointly model signal transduction and gene expression, these stochastic promoter models were coupled to deterministic models of signaling, and this yielded insights into the dynamics of target gene expression [97, 98] and into the long-term regulation of signaling heterogeneity by stochastic signaling protein expression fluctuations [82].

In certain cases, intrinsic stochastic dynamics may arise within the signaling network, especialy if the pathway operates at low molecule numbers [51, 98, 99]. The level of initial signal sensing may be especially prone to stochastic dynamics, since cell surface receptors are often only expressed at a few hundreds or thousand molecules per cell [52, 100]. Then, processes like receptor endocytosis which simultaneously remove hundreds of molecules at once from the cell surface, may give rise to digital behavior and consequently strong stochastic fluctuations of signaling activity [101]. Accordingly, stochastic models of signaling networks have been proposed to describe heterogeneous decision making [51, 97, 99, 100], and several of these studies focussed on fluctuations at the receptor level [97, 99, 100].

### 2.4 Quantitative modeling of cellular heterogeneity

In many cases, ensemble modeling approaches are semi-quantitative in the sense that the kinetic reaction parameters and the protein fluctuations (i.e., the standard deviation of their distribution) are tuned manually. While such semi-quantitative modeling is valuable in many cases, the long-term goal is a quantitative match between model and experiment. This can be achieved by directly fitting the single-cell models to single-cell data by minimizing the difference between the simulated and measured single-cell distributions.

Quantitative singe-cell model fitting of signaling and gene expression has now been applied in a number of publications and is an lively area of research [57, 60, 85, 86, 88–90, 94, 102–111]. In several of these studies, the fitted models involve deterministic sources of heterogeneity [85, 86, 107], stochastic fluctuations [94], or a combination of both [60, 88, 105]. In the deterministic case, only signaling protein concentrations may be cell-specific parameters [85] or all model parameters may show cell-to-cell variability [89]. For a comprehensive overview over the assumptions and the computational methods, we refer to the recent review by Hasenauer and Loos [112].

The methods for model calibration can be classified based on the type of experimental data they use, single-cell snapshot or time course data (Fig. 2D and [112]). For snapshot data, only distributions at single time points are considered and therefore potential correlations between observations at consecutive time points are neglected. Despite this limitation, the approach benefits from the fact that snapshot data can typically be generated on a higher throughput, i.e., for more cells and molecular species, when compared to time-resolved measurements. For instance, high-throughput snapshot data can be generated using flow cytometry, mass cytometry or single-cell RNA sequencing, and the larger wealth of information should be beneficial for the training of reliable mathematical models. Accordingly, several approaches were were proposed for the model calibration based on snapshot data [88–90, 102, 104, 111]. In deterministic models, different assumptions were made about the type of fluctuations in parameter values, ranging from the unimodal log-normal distribution of to multimodal distributions, or even no specific assumption was made about the nature of fluctuations (non-parametric distribution; reviewed in [112]). For instance, Hasenauer et al. employed multi-modal mixtures of normal parameter distributions to infer subpopulations (with distinct mean parameters) by fitting a model NGF signaling to snapshot data [102].

A limitation of the snapshot approach is that essential information about temporal behavior in single cells gained from live-cell imaging (e.g., an oscillatory pattern) may be lost. Therefore, snapshot information may be less well suited for the identification of molecular sources of heterogeneity when compared to time-resolved data (discussed in [110]). As a consequence, several studies suggested to directly fit a mechanistic model to single-cell time course data [85, 86, 103, 105–110]. In naive approach, each individual cell could be fitting separately by minimizing the residuals between model and data, and then the single-cell fitting results are combined to yield cell population distributions of interest, e.g., for signaling protein expression levels. In this so-called standard two-stage approach, each cell is thus analyzed as a independent sub-problem (stage 1), and then the cell population is assembled (stage 2). However, stage 1 suffers from the problem that the model parameters in systems biology models can almost never be correctly estimated (identified), especially based on live-cell imaging data which typically only covers one molecular species for each cell. Thus, the fitting uncertainties are high and the heterogeneity between cells is overestimated [106, 107] which limits the predictive power of the two-stage approach unless very small models are considered.

To circumvent this problem, information about the cell population distribution needs to be taken into account during fitting of single-cells [85, 106, 107, 110]. Specifically, the fitted likelihood function combines information from all cells, and these additional constraints improve the identifiability of single-cell parameters which leads to smaller uncertainties in model predictions. For instance, as a constraint in deterministic models, it can be ensured that protein concentration fluctuations follow a log-normal distribution [85, 106]. Moreover, the model can be simultaneously fitted to single-cell and population-average data, and it has been shown that the combination of both types leads to a better discrimination of model variants compared to the use of either alone [85].

It should be noted that the current quantitative models of cellular signaling and gene expression heterogeneity are typically limited to a few species and reactions. Therefore, qualitative ensemble modeling approaches (Section 2.2) are still very valuable for large-scale networks and typically led to more profound “biological” insights when compared to quantitative approaches. Further improvements are needed in the computational methods for quantitative model fitting to reduce computational cost and to integrate various types of data including cross-sectional snapshots, high-resolution live-cell imaging and population-average data. This will improve identifiability of model parameters, the certainty of model predictions and will be helpful to discriminate competing model variants also in larger networks.

## 3. Heterogeneity in TGFβ signaling - modeling and impact on cellular behaviour

In the final part of the chapter, we discuss the characteristics and modeling of TGFβ/SMAD signaling at the single-cell level. We initially start with an overview over the pathway and its role in controlling cell fates. Then, we summarize its dynamic features at the single-cell level and outline how population-average as well as single-cell modeling approaches provided insights into the pathway dynamics. Finally, we review recent work, in which the link between fluctuations in SMAD proteins and target gene expression was explored.

### 3.1 TGFβ signaling in health and disease

TGFβ belongs to a family of soluble extracellular ligands that activate intracellular signaling by binding to cell surface receptors. As depicted in Fig. 3A, signaling is initiated by a cascade of events that involves TGFβ binding to the TGFβR2 receptor, and this receptor-ligand complex in turn binds to the TGFβR1 receptors (also known as ALK5) to build an activated receptor complex [113] . The active TGFβ receptor functions as as a intracellular kinase that phosphorylates cytoplasmic SMAD2/3 proteins which upon phosphorylation form heterotrimers with SMAD4 (e.g., $(SMAD2)_2(SMAD4)$). SMAD heterotrimers translocate to the nucleus and there act as transcription factors, i.e., they bind to and activate gene promoters to regulate the target gene expression [114]. The signaling pathway activity is terminated by nuclear dephosphorlyation of SMAD proteins, dissociation of the complexes and finally the nuclear exit of SMAD proteins.

TGFβ induces several cellular responses including cell cycle arrest, apoptosis and cell migration [115]. TGFβ-induced cell migration typically involves the so-called epithelial-to-mesenchymal transition (EMT), a phenotypic remodeling of cells in which the cytoskeletal reorganization and loss of cell-cell junctions allows epithelial cells to evade from their original location by acquiring a motile, migratory, mesenchymal phenotype [116]. Given these widespread roles in cellular remodeling, it is not surprising that TGFβ and closely related ligands (e.g., GDF11 or BMPs) play a critical role in embryogenesis and tissue homeostasis, but also in diseases such as cancer or fibrosis [117]. For instance, in higher vertebrate development, gastrulation and neural crest formation depend on EMT induced by TGFβ superfamily members [116] . Likewise, TGFβ induced apoptosis and cell cycle arrest maintain tissue homeostasis and prevent overgrowth in developing and adult tissues, e.g., in the liver [118]. If the cytostatic effect of TGFβ or is its ability to induce apoptosis is lost this leads to tumor progression. However, TGFβ signaling not only acts as a tumor suppressor, but plays a dual role in cancer progression, as in late-stage tumors aberrant TGFβ induced EMT signaling promotes the formation of metastasis [119]. Thus, during cancer development, a specificity switch occurs, in which TGFβ signaling no longer promotes cytostatic responses, but mostly induces cell migration.

At the molecular level, this specificity switch involves a change in the set of target genes regulated by TGFβ/SMAD signaling: In late-stage tumors, TGFβ no longer downregulates growth-promoting oncogenes like Myc and fails to upregulate cell cycle inhibitors (e.g., p15, p21). Instead, TGFβ induces gene expression changes that are crucial for mediating early steps of reprogramming from epithelial to mesenchymal identity including the downregulation of classical epithelial and upregulation of mesenchymal markers [119]. Based on experimental evidence, several hypotheses have been been proposed to explain how the same signaling pathway can induce qualitatively distinct gene expression responses depending on the cellular context: (1) context-specific expression of transcription co-factors involved in SMAD-dependent gene expression [120], (2) alterations in the concentrations of SMAD2 and SMAD3 each of which controls specific sets of target genes [121] or (3) encoding of specific gene expression programs by the temporal dynamics of the SMAD pathway [122, 123]. Specifically, it has been suggested that a transient SMAD signal may be sufficient for EMT and cell migration, while sustained signaling additionally triggers cell cycle arrest [122]. Thus, quantitative insights into the pathway dynamics by time-resolved live-cell imaging [18] and mathematical modeling are important to better understand cellular responses to TGFβ stimulation.

### 3.2 Lessons learned from single-cell experiments of TGFβ signaling

At the single-cell level, individual cells respond very differently to TGFβ treatment. Due to this heterogeneity, time-resolved analyses of SMAD signaling at the single-cell level are valuable tools to understand the link between signaling dynamics, gene expression and cellular outcome. In fact, single-cell studies further supported that the amplitude and/or duration of the SMAD signal partially determines whether a cell will react at all to stimulation and/or whether it will respond with migration or cell cycle arrest [33, 36].

Established experimental readouts of TGFβ/SMAD signaling at the single-cell level include measurements of receptor levels and their internalization [124], nuclear translocation of SMAD proteins [33, 34, 41, 125, 126], SMAD trimerization [127] and SMAD-induced gene expression [34, 72]. At the signaling level, SMAD nuclear translocation assays are most widely used. They rely on the mild overexpression of SMAD2 or SMAD4 fluorescent fusion proteins, and the nuclear translocation of the fluorophore is then used as a proxy for pathway activation. [33, 34, 36, 41, 126, 128] By automated microscopy, images are taken on a temporal resolution of a few minutes. Subsequently image analysis is performed to track cells, and to segment them into nuclear and cytoplasmic compartments. The amount of SMAD-associated fluorophore in nucleus or the nucleus-to-cytoplasmic fluorescence ratio is then used as a measure of signaling pathway activity. Depending on the study, this technique allowed for signaling analysis in 250-1500 single cells per condition over a time frame of 45 min to 24h.

These large-scale single-cell datasets indicated that heterogeneous SMAD signaling at the single-cell level exhibits a few key features that are recurrently observed across experimental groups and cellular systems:

***1) SMAD signaling is gradual at the single-cell level (Fig. 3C):*** In Section 1, it was discussed that certain signaling pathways like the MAPK and apoptosis cascades, show bimodal behavior, i.e., complete or no activation in individual cells. Available single-cell studies on TGFβ signaling suggest that this pathway rather acts like a gradual continuum, i.e., snapshot histograms of SMAD2 nuclear translocation show a unimodal distribution with strong cell-to-cell variability (Fig. 3C). With increasing TGFβ doses, this mean value of this continuous distribution gradually shifts to higher signaling levels [33, 34, 41, 125, 128]

***2) Single cells show qualitative differences in signal shape (Fig. 3B):*** Single-cell analyses show strong differences in the absolute level of SMAD2 or SMAD4 nuclear translocation between cells at a given time point [33, 34, 36, 41, 128]. In time-resolved analyses, individual cells may also be distinct in the shape of the signal, e.g., in the kinetics or the degree of adaptation to a lower pleateau after the initial peak amplitude. Such heterogeneity in the shape of the signal was observed either at the level of SMAD2 or SMAD4 nuclear translocation [33, 34, 125], or at the level of target gene expression [34]. In contrast, in other cellular systems the shape of SMAD2 nuclear translocation was fairly similar between cells [129]. Clustering techniques using dynamic time warping as a similarity measure between cells were used to sort the trajectories of individual cells into classes with qualitatively different dynamics [33, 34]. Using this clustering approach, we found that even for a given TGFβ dose some cells do not respond to the stimulus (non-responders), others show a transient response, whereas the remainder show sustained pathway activation (Fig. 3B). At very low TGFβ stimulation, most cells belong to the non-responding cluster, whereas at intermediate and high TGFβ doses, the transient and sustained clusters predominate, respectively. Population-average measurements are a mixture of these qualitatively distinct responses and therefore only partially cover the complexity of the pathway at the single-cell level. Interestingly, the signalling clusters are better predictors for TGFβ-induced cell migration and division when compared to the applied extracellular ligand dose [33]. This further suggests that the cellular decisions are linked to SMAD signaling dynamics at the single-cell level.

***3) Single cells show burst-like shuttling of SMAD proteins (Fig. 3B):*** Upon stimulation, the population-average response of the SMAD signaling pathway typically shows an initial peak amplitude ~60 mins after stimulation. Afterwards, the population-average signal slowly declines over a time scale of several hours, but may remain constantly elevated, e.g., upon strong TGFβ stimulation, but this depends on the cellular context. For such sustained behavior, the single-cell response is distinct from the population average and shows repeated bursts of nuclear translocation (Fig. 3B). Specifically, the nuclear SMAD2 or SMAD4 levels decline strongly after the initial peak, before again reaching once or multiple times levels comparable to the level of the initial peak [33, 34]. This did not appear to a technical artefact of imaging, as a generic nuclear marker (H2B) did not show bursting behavior [130]. Furthermore, SMAD4 bursts were reported in developing Xenopus embryos in which TGFβ family ligands play an important role, and the behavior could be reproduced in isolated animal cap explants [128] Interestingly, these pulsatile SMAD translocation dynamics are irregular in their timing intervals and amplitudes, suggesting that they may, in part, arise from stochastic dynamics of the SMAD signaling pathway.

In the following, we will discuss how mathematical models can provide insights into these dynamical features at the single-cell level. We will first review population-average models of TGFβ/SMAD signaling and will then turn to modeling approaches at the single-cell level.

### 3.3 Population-average models of TGFβ/SMAD signaling

Early kinetic models of TGFβ/SMAD signaling mainly focused on the description of Western Blot measurements of SMAD2/3 phosphorylation and complex formation [131–139]. Since these experimental methods provide average quantifications of thousands to millions of cells, the resulting models describe the behavior of a representative average cell and fail to capture heterogeneity in the population. Population-average models of SMAD signaling are typically based on deterministic ordinary differential equations (ODEs), and the individual reaction steps are formulated based on mass-action kinetics. Models proposed in literature have been reviewed elsewhere [140, 141] and differ in the level of detail they consider and in the reaction mechanisms they focus on. Still, most of the models share a set of key mechanisms including receptor-ligand binding, receptor shuttling to the endosome, receptor-mediated SMAD phosphorylation SMAD (de)phosphorylation, trimerization and nuclear translocation, as well as transcriptional negative feedback via target genes that, for instance, inhibit receptor signaling (Fig. 3A). The kinetic parameters are typically not known and were estimated by fitting the models to experimental data [33, 135, 142], or the parameter space was explored by random sampling or sensitivity analysis [132, 134, 143, 144].

Interestingly, population-average models alongside with quantitative experiments reproduced and provided insights into several features of heterogeneous TGFβ/SMAD signaling at the single-cell level including the gradual behavior of the pathway, transient vs. sustained signaling and pulsatile pathway dynamics. Thus, they provide hints to mechanisms of heterogeneity and serve as a basis for deterministic modeling of variability at the single-cell level.

***1) Gradual dose-response behavior:*** Population-average modeling studies and dose-response measurements revealed gradually increasing SMAD signaling in response to increasing doses of TGFβ. Specifically, intracellular SMAD signaling exhibits a shallow dose-response curve with a Hill coefficient ($n_H$) of close to or less than 1 [33, 72, 136, 145]. Modeling studies further revealed that switch-like dose-response behavior ($n_H$=4.5) late after stimulation (24 h) is *not* an inherent feature of the SMAD signaling pathway, but arises from degradation of extracellular TGFβ in the cell culture dish [136, 146]. Hence, the SMAD signaling pathway models respond gradually to perturbations (in both TGFβ concentration and intracellular protein concentrations) and are therefore consistent with a unimodal SMAD nuclear translocation distribution at the single-cell level.

***2) Features controlling signal amplitude and duration:*** Sensitivity analysis of population-average models revealed the relative importance of individual reaction steps in controlling the signal amplitude and duration [133–136]. It seems that the signal duration (i.e., the shape of the signal) is mainly set by the kinetics of receptor-ligand binding and receptor shuttling. Accordingly, SMAD nuclear translocation cycle typically shows similar dynamics as the receptor level and mainly acts as a remote sensor that directly reflects receptor changes, though with a slight time delay of ~10 minutes [133, 147]. Molecular mechanisms that control the signaling dynamics at the receptor (and thus the SMAD) level include: (i) Receptor downregulation from the cell surface by internalization of receptor-ligand complexes into endosomal compartments [137, 148]; (ii) Cell-mediated degradation of extracellular TGFβ, again by internalization of receptor-ligand complexes and subsequent intracellular degradation of the ligand. [131, 136]. This mechanism of signal termination becomes an important factor in controlling the length of the signal if the number of extracelluar TGFβ molecules per cell is limiting (low TGFβ concentration and/or small extracellular medium volume) (iii) Negative feedback of SMAD target genes to the receptor level, e.g., by SMAD-induced expression of inhibitory SMAD7 and BAMBI proteins. These proteins bind to TGFβ receptor complexes, thereby inhibiting their kinase activity and targeting them for degradation [33, 134, 139]. Taken together, the signal duration is controlled by multiple mechanisms at the receptor level. Using sensitivity analysis of population-average models, a similar multi-level regulation by many reaction steps in the pathway was shown for the absolute scale (i.e., the amplitude) of the SMAD signal [133, 135]. At the single-cell level, such mechanisms jointly control heterogeneous signaling dynamics and this can be investigated by parameter sampling in a deterministic model (see below).

***3) Pulsatile SMAD shuttling*:** Population-average modeling and quantitative experimental analyses suggested that SMAD signaling induced by TGFβ or BMP could show (damped) oscillatory behavior, in which a single stimulus induces two or more repeated pulses of SMAD nuclear translocation [134, 148]. Using global parameter sampling, Wegner et al proved that oscillations require the presence of transcriptional negative feedback - if this feedback is switched off, no physiologically plausible parameter configuration can produce oscillations [134]. Accordingly, knockdown of transcriptional feedback regulators SMAD6 and SMAD7 abolished BMP-induced SMAD oscillations [148]. It is possible that such oscillatory negative feedback contributes to repeated bursting of SMAD2 or SMAD4 nuclear translocation observed in single cells, though additional stochastic mechanisms need to be taken into account to describe the irregularity of bursts [33, 128].

### 3.3 Towards quantitative modeling of SMAD signaling heterogeneity

On the basis of the established population-average models, we recently derived a deterministic modeling framework to quantitatively describe cell-to-cell variability in the TGFβ/SMAD signaling pathway [33]. Our study was based on imaging data in which the nuclear translocation of SMAD2/4-GFP fusion proteins was monitored in thousands of living MCF10A cells over 24 h.

As a basis for modeling, we initially analyzed characteristic features of SMAD signaling heterogeneity. We performed sister cell experiments and found that sister cells are more similar than random pairs of cells but desynchronize after several hours. This indicated that signaling fluctuations are non-genetic, but temporally stable. We then analyzed potential sources of heterogeneity in SMAD signaling, and considered that SMAD signaling may be influenced by cell cycle stage and/or cell density [127, 149]. Using live-cell imaging, we followed cell division events and quantified the cell density before and during TGFβ stimulation, and found that these two factors had negligible impact on heterogeneous signaling in our culture conditions. Taken together, this indicated that SMAD signaling heterogeneity can be modeled using a deterministic approach based on ODEs (Section 2), with the assumption of stochastic (but temporally stable) fluctuations in signaling protein expression levels.

We started with a detailed kinetic pathway model that describes known mechanisms of SMAD signaling and comprises a total of 23 molecular species and 45 kinetic reaction parameters. Like most other models of TGFβ signaling, our model contained the main features described in Section 3.3, including an endosomal receptor shuttling module, a SMAD translocation module and a transcriptional feedback module (Fig. 3A). With this model, we sought to describe a large experimental dataset, in which several levels of signaling (TGFβ receptor expression, nuclear translocation of SMAD2 and SMAD4, as well as SMAD7 mRNA expression) were measured for multiple experimental conditions at the single-cell and population-average levels. In total, this dataset comprised >1,000,000 data points, mainly from densely sampled live-cell imaging experiments at multiple experimental conditions. Owing to the high complexity of model and data, we did not aim for a quantitative fitting of the model to the single-cell data, but instead devised a three-tiered, modeling strategy to derive a quantitative description of heterogeneous signaling (Fig. 3D). Initially, we describe the population-average dynamics. Then, we refine the description of the pathway to the level cellular sub-populations showing qualitatively distinct signaling dynamics. Finally, we develop an ensemble of single-cell models to describe the complete heterogeneous cell population. Specifically, the three modeling steps were as follows:

***1) Population-average modeling*:** To derive a quantitative description of the SMAD signaling dynamics, we initially fitted the model to population-average data (Fig. 3D, left). For fitting, we used the population-median nuclear translocation time courses of fluorescently labeled SMAD2 and SMAD4 for varying TGFβ doses and for restimulation experiments, in which cells were repeatedly challenged with the ligand. Furthermore, the fitting took into account pathway measurements that were only possible at the population-average level (TGFβ receptor protein expression, SMAD7 mRNA expression). After calibration, we validated the predictive power of our model for previously untested molecular species (time course of extracellular TGFβ degradation) and experimental conditions (restimulation experiments, inhibition of transcriptional feedback loops by small molecule inhibitor DRB).

***2) Description of cellular subpopulations*:** Having a predictive population-average model at hand, we sought to quantitatively describe variability in signaling, while limiting the computational cost. Therefore, we refrained from fitting our model to the complete single-cell population (Section 2.4), but only fitted six subpopulations which show qualitatively distinct dynamics of signaling (e.g., transient vs. sustained; see Fig. 3D, middle). These subpopulations were identified by k-means clustering of single-cell SMAD2 nuclear translocation time courses according to their similarity in shape and amplitude. We separately fitted the subpopulation-median time course of each cluster and only allowed variation in the expression of signaling proteins (e.g., TGFβ receptors, SMADs) within the range of typical cell-to-cell variation (+/- 2-fold). In contrast, the kinetic parameters were fixed to their population-average value, i.e., their variability was neglected. With these assumptions, we could quantitatively describe all subpopulations, and had therefore developed our model from a population-average description to a description of six representative cells with characteristic dynamical features.

***3) Ensemble modeling of complete cell population:*** To directly compare our simulations to single-cell experiments, we converted the subpopulation models to an ensemble of artificial cells representing the heterogeneity of the entire cell population (Fig. 3D, right). Artificial single cells belonging to each subpopulation were generated by repeated simulation with signaling protein concentrations varying around the best-fit values of the corresponding subpopulation model. The full cell population was assembled *in silico* by combining artificial cells according to the experimentally observed proportion of corresponding subpopulations. The degree of variation was assumed to be the same for all sampled signaling proteins. The common protein coefficient of variation (std/mean) was chosen by matching the simulated and and experimentally observed snapshot distributions at particular time points using summary statistics.

Taken together, we obtained an *in silico* cell population with realistic properties close to the experimental data. Importantly, we could show that our three-tiered modeling approach, in which we considered the subpopulation structure (step 2), yielded a better agreement with single-cell snapshot distributions (step 3) when compared to direct sampling of protein concentrations in the population-average model. The model reproduced key features of single-cell TGFβ signaling including gradual (unimodal) behavior and strong heterogeneity in the time course shape (Section 3.2). By calculating euclidean distances, we quantitatively compared the simulated single-cell trajectories to the six experimentally observed time course clusters. Since the model took into account subpopulation information, we obtained a very similar dose-dependent decomposition into non-responding, transient and sustained signaling classes as for the experimental data. Thus, the model correctly takes into account temporal correlations in signaling pathway activity, and can be used to predict drifts in the shape and proportion of the original subpopulations for any experimental condition. In fact, we confirmed such predictions to a knockout of the negative feedback regulator SMAD7. As predicted by the model, we found that the effect of SMAD7 on the signaling dynamics was restricted to certain cellular subpopulations and was observed for specific doses of TGFβ only. Hence, the model allowed us to quantitatively understand the cell-specific impact of experimental perturbations and allowed mechanistic insights into cellular heterogeneity.

One limitation of the current model is that the best-fit parameter values in population-average and subpopulation fitting are not unique (non-identifiability problem). Nevertheless, robust predictions could be made, since very similar simulation results were obtained when comparing multiple fits obtained during a multi-start optimization (repeated model fitting from different starting parameters). The subpopulation fitting (step 2) currently corresponds to a the standard two-stage approach discussed in Section 2.3, since the protein concentrations in each subpopulation were estimated separately without additional constraints about the protein distribution in the cell population. After assembly of the complete cell population (step 3), we confirmed that signaling protein levels in the model show a realistic log-normal distribution. However, such a distribution is not automatically granted in the current approach. Therefore, it would be beneficial to either improve the identifiability of parameters by model reduction, or to take into account additional constraints during subpopulation fitting.

Taken together, our study suggest that heterogeneity of TGFβ/SMAD signaling at the single-cell level can be quantitatively described using a deterministic modeling approach. Key features of the pathway at the single-cell level were reproduced, including gradual behavior and cell-specific characteristics in signaling shape. Since the model is deterministic in nature, it currently does not describe the apparently stochastic, burst-like shuttling of SMAD proteins into the nucleus (Section 3.2). To describe this phenomenon, stochastic effects, most likely in endosomal receptor shuttling need to be taken into account, and quantitative fitting approaches could be used to match burst features in  the stochastic model to the experimental data. This may be an interesting future direction, as other signaling pathways such as the MAPK cascade show repeated pulses of activation which have profound impact on cell fate [43, 150].

### 3.4 Future directions - Link between signaling and gene expression heterogeneity

Another aspect that deserves further attention in the future is how fluctuations in signaling proteins translate into fluctuations in downstream target gene expression. SMAD transcription factors mediate cellular responses by binding to target gene promoters, thereby inducing large-scale gene expression programs involved in cell migration, EMT and cell cycle arrest [115]. Given this link between SMAD binding and gene expression, it seems that cell-specific gene expression and morphological changes may be predictable based on the amplitude and/or dynamics of SMAD signaling. In fact, by clustering single-cell SMAD time courses, we found that cell migration and cell division kinetics can - in part - be explained based on the dynamics of SMAD nuclear translocation [33].

Based on these observations, it is natural to extend current mathematical models to SMAD-induced gene expression and -in the long run– to TGFβ-induced cell fate decisions. At the population-average level, SMAD signaling dynamics seem indeed to be related to gene expression, as models simultaneously describing time courses at both levels are well-established: most of these studies modeled the dynamics of certain negative feedback regulators or pathway targets using SMAD-dependent transcription and linear degradation of the target gene. By simultaneously fitting such synthesis and decay models using SMAD kinetics as an input, target gene mRNA dynamics can be well-explained in multiple cases [33, 72, 134, 143, 146, 151, 152]. A recent combined modeling and experimental study extended this idea to a larger set of target genes based quantitative and time-resolved measurements SMAD trimeric complexes [142]. Using a model fitting framework, the authors inferred the specificity of target gene induction by distinct SMAD complexes and successfully predicted gene expression outcomes for novel experimental  conditions.

Despite such good accordance of SMAD signaling and gene expression responses in cell populations, it remains challenging to directly link both levels in single cells. Experimentally, such analyses require SMAD signaling dynamics and gene expression to be simultaneously measured in the same cell by live-cell imaging of fluorescent SMAD reporters and subsequent smFISH of mRNA expression in fixed cells [34, 41]. Those two studies conducted so far, agree that on a single cell level, neither SMAD2 nor SMAD4 absolute levels in the nucleus accurately predict the cell-specific expression of target mRNAs. However, Frick et al. reported that the TGFβ-induced fold-change of the nuclear SMAD levels relative to basal predicts stimulus-induced target gene expression responses. For the genes Snail and CTGF, they found Spearman rank correlation coefficients of ~0.5 between between those fold-changes and the mRNA abundance as measured by smFISH. These findings could not be confirmed by Tidin et (2019), who analyzed the expession of CTGF with high temporal resolution using live luminescence imaging. They found no correlation between the fold-changes in nuclear SMAD and CTGF expression. Thus, even though the proposed fold-change detection in SMAD target gene expression may be an elegant way to reliably respond to stimulation despite high variability in absolute nuclear SMAD levels, it remains to be confirmed whether this is general phenomenon applicable to other genes and cellular systems.

Therefore, the mechanistic link between SMAD signaling fluctuations, gene expression and heterogeneous cellular responses remains to be established. In any case, a deterministic 1:1 correspondence of signaling and gene expression appears unlikely, since gene expression modeling requires stochastic modeling of promoter switching, as Molina et al showed for SMAD target genes (see also Section 2.3) [153]. Even though each individual gene might respond stochastically and with little correlation to the SMAD signal, it still remains possible that SMAD signaling fluctuations directly affect cellular outcomes through their cumulative effect on many target genes controlling a common biological process. Genome-wide single-cell RNA sequencing approaches will shed light on such coordinated gene expression programs at the level of individual cells.

Recent work has presented methods for the targeted manipulation of SMAD signaling dynamics at the single-cell level [36, 126] and similar tools were developed for other signaling pathways [43, 83, 154]. The combination of such highly controlable tools with gene expression measurements will provide direct insights into the impact of SMAD signaling dynamics on gene expression outcomes, and will advance our understanding of heterogeneous decision making by the TGFβ pathway.

# FIGURE LEGENDS

**Figure 1: Functional and physiological consequences of non-genetic heterogeneity.**

(A) Heterogeneous signaling causes fractional cell differentiation. Only cells exhibiting high signaling activity (dark grey) in response to a differentiation stimulus undergo differentiation, e.g., during MAPK-induced oocyte maturation or PC12 differentiation (see text).

(B) Signaling variability in response to therapeutic treatment. Apoptotic signaling pathways (e.g., caspases) are activated heterogeneously when cells are treated with a cytotoxic drug which results in fractional killing and (transient) resistance in the non-responding cells. A population regrown from the therapy-resistant cells may again exhibit the same fractional killing, indicating that signaling heterogeneity is a non-genetic phenomenon.

(C) Heterogeneous signaling may exhibit gradual or bimodal behavior. Gradual signaling pathways exhibit a unimodal activity distribution across single cells which is shifted to higher mean levels upon increasing stimulation. Signaling histograms at different doses typically overlap, which may give rise to inaccurate cellular information transfer. Signaling systems with bimodal behavior exhibit two clearly separable (“ON” and “OFF”) activity levels. Increasing stimulation does not affect the mean signaling activity of the ON and OFF subpopulation, but shifts the fraction of cells in each class.

(D) Sister cells experiments indicate deterministic behavior of signaling. For a signaling pathway with stochastic dynamics, even recently divided sister cells would show distinct (stochastic) signaling responses. In signaling, recently divided sister cells are typically more correlated in their signal response than random cells, likely because sisters show common protein expression patterns or cell cycle stages. Over a timescale of hours to days, sister cell similarity is lost (“older sister cells”), indicating a non-genetic mechanism of signaling heterogeneity.

(E) Variations in signaling protein expression cause deterministic heterogeneity of signaling. A high total expression level of a positive regulator ($X_{tot} = X + X^*$) causes a strong response in a signaling pathway with deterministic behavior. In a signaling pathway with stochastic dynamics, the signaling response is less or not at all related to the protein content. In models of cellular heterogeneity, protein expression is often assumed to be stable at the time scale of signaling.

**Figure 2: Deterministic modeling of cellular heterogeneity.**

(A) A kinetic model of securin and cyclin B degradation by the anaphase-promoting complex (APC) during mitosis (left) is described by a deterministic ODE model (see also [75]) and numerically integrated to simulate the protein dynamics in one average cell (middle). Heterogeneity is introduced into the system by performing repeated simulations while sampling protein concentrations (and in some cases kinetic parameters) from log-normal distributions (right).

(B) Model-based analysis of cellular heterogeneity. For model analysis, simulations are performed for varying parameters and/or degree of protein concentration fluctuations, or different biological mechanisms are considered in the model. Thereby, the model provides insights into molecular sources of heterogeneity, mechanisms of biological robustness, and allows for the design of new experiments.

(C) Quantitative fitting of ensemble models to heterogeneous single-cell data. In the literature, models of cellular heterogeneity were calibrated by fitting cross-sectional snapshot data at a particular time point, or by directly fitting the kinetic model to single-cell trajectories. In both cases, cell-specific parameters are estimated to yield an optimal match between model and experiment.

**Figure 3: Quantitative analysis and modeling of TGFβ/SMAD signaling heterogeneity**

(A) Schematic representation of TGFβ signaling including endosomal receptor shuttling, nuclear SMAD shuttling and negative feedback: The binding of the ligand to a RII receptor leads to the recruitment of an RI receptor building an activated receptor complex. This complex can, when internalized, mediate the phosphorylation of SMAD2 proteins. The receptor complex either gets degraded or free receptors are recycled back to the cells surface. In the cytosol, phosphorylated SMAD2 proteins form trimers with SMAD4 which translocate to the nucleus, thereby incrreasing the experimentally measurable nuclear-to-cytoplasmic (Nuc/Cyt) SMAD2 ratio. In the nucleus, the SMAD heterotrimer acts as a transcription factor to induce downstream target genes of TGFβ including SMAD7 which acts as a negative feedback inhibiting the activity of RII receptors.

(B) Temporal dynamics of SMAD nuclear translocation at the single-cell level. Population average (black) and standard deviation (gray shades) of the single-cell Nuc/Cyt SMAD2 ratio after stimulation with 100 pM of TGFβ. Examplary trajectories of transient, sustained and non-responding cells are shown in different colors (see legend). Another cell (purple) shows a bursting event (red). Data from [130].

(C) Gradual behavior of SMAD signaling at the single-cell level. Snapshot histograms of Nuc/Cyt SMAD2 ratio 70 min (time of peak) after stimulation with varying doses of TGF-b (see legend). All distributions are unimodal and shift towards higher mean values with increasing stimulation. Data from [130].

(D) Three-tiered modeling approach for modeling TGFβ/SMAD signaling heterogeneity: In step (1), a population-average model is fitted to experimental data. In step (2), the model is refined to a desription of subpopulations which are identified from the measured single-cell trajectories using a clustering approach. For each subpopulation, the model is fitted to the median time courses of a cluster, assuming subpopulation-specific signaling protein expression. Step (3) yields simulations of individual cells, since signaling protein levels are sampled from log-normal distributions in each subpopulation model. In step (4) single-cell trajectories from step (3) are combined according to yield a description of the complete cell population. Figure modified from [130]

# FIGURES

## Figure 1

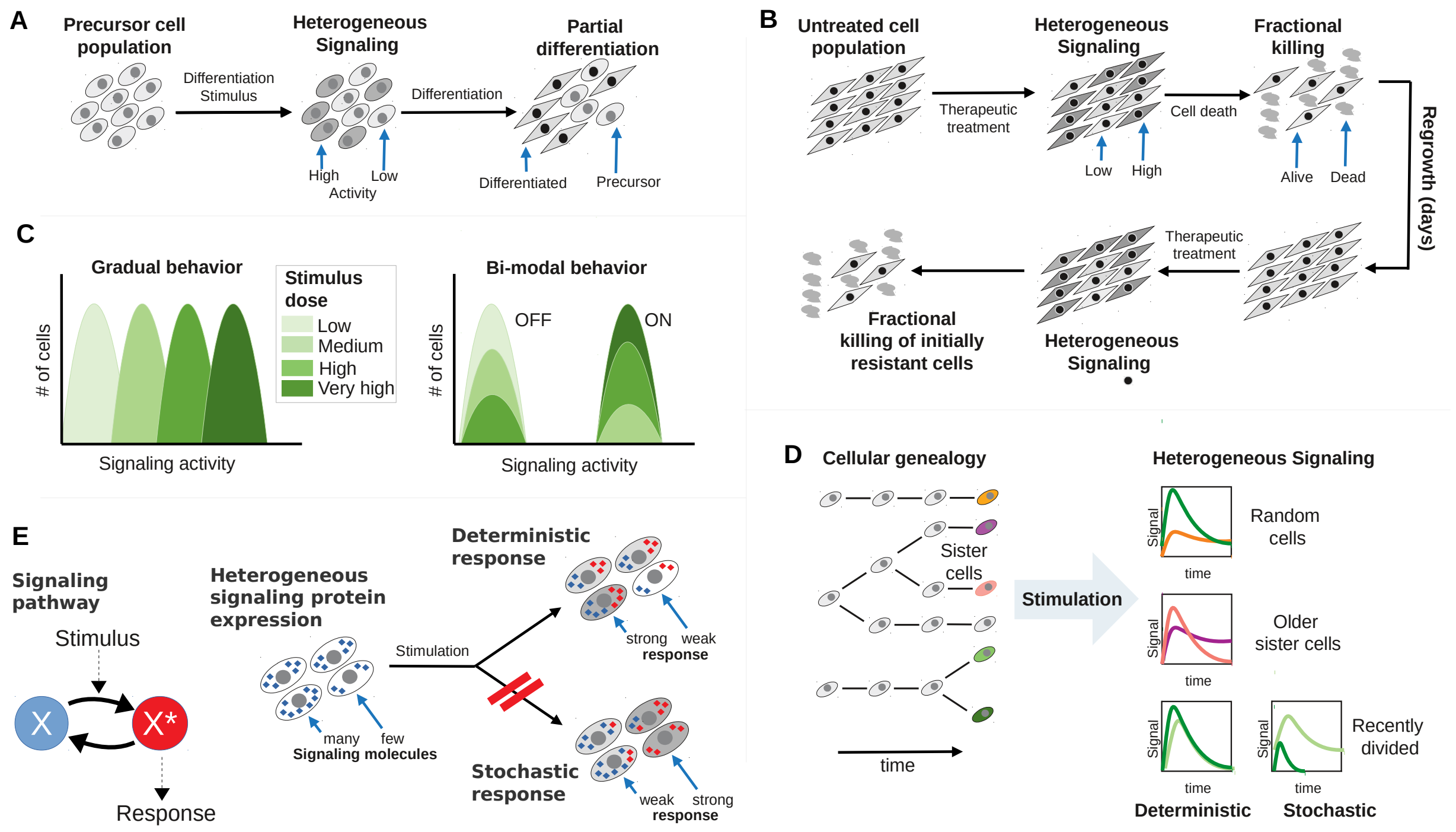

## Figure 2

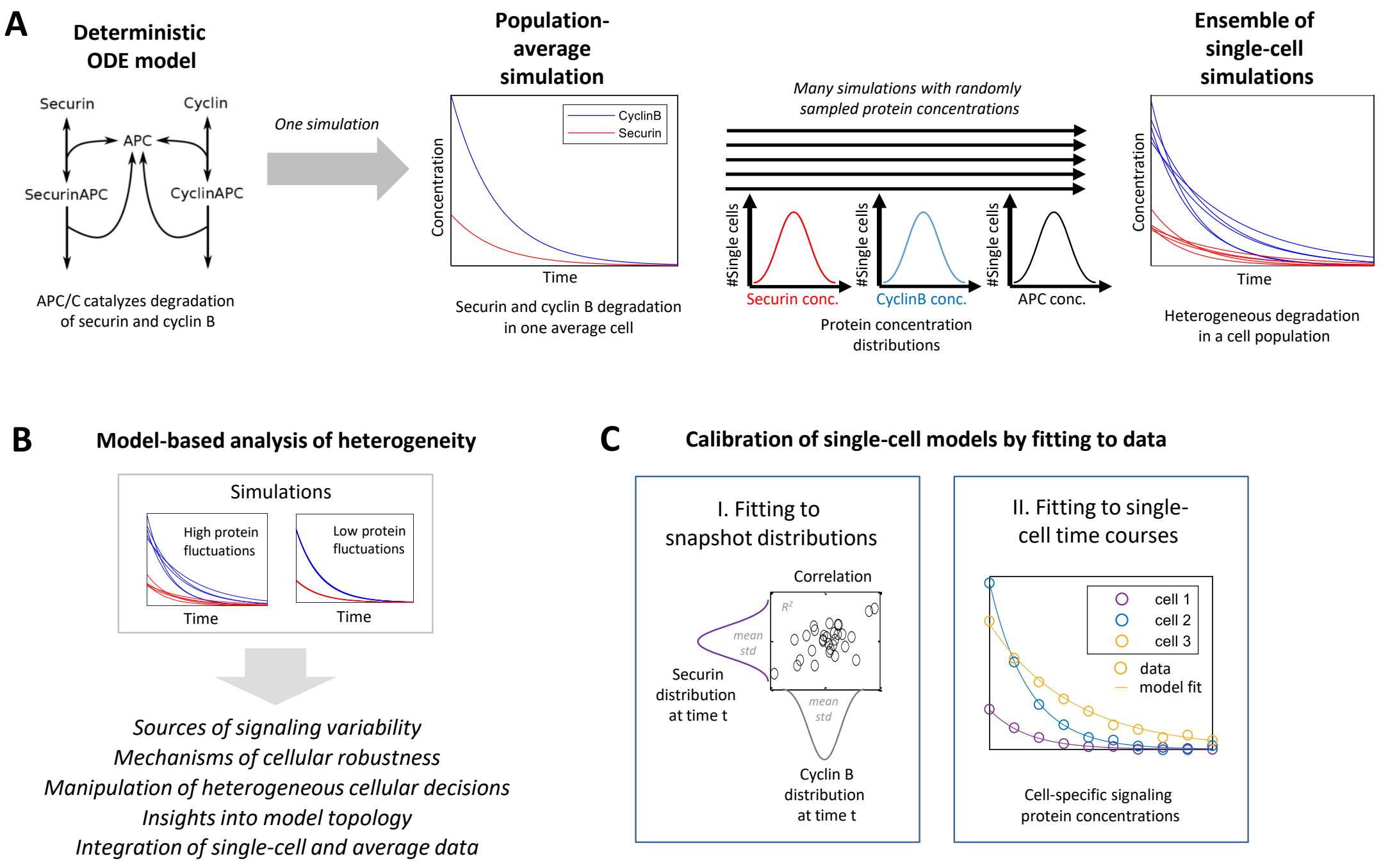

A
Deterministic ODE model
Securin
Cyclin
APC
SecurinAPC
CyclinAPC
APC/C catalyzes degradation of securin and cyclin B
One simulation
Population-average simulation
CyclinB
Securin
Concentration
Time
Securin and cyclin B degradation in one average cell
Many simulations with randomly sampled protein concentrations
#Single cells
Securin conc.
CyclinB conc.
APC conc.
Protein concentration distributions
Ensemble of single-cell simulations
Heterogeneous degradation in a cell population
B
Model-based analysis of heterogeneity
Simulations
High protein fluctuations
Low protein fluctuations
Sources of signaling variability
Mechanisms of cellular robustness
Manipulation of heterogeneous cellular decisions
Insights into model topology
Integration of single-cell and average data
C
Calibration of single-cell models by fitting to data
I. Fitting to snapshot distributions
Correlation
mean
std
Securin distribution at time t
Cyclin B distribution at time t
II. Fitting to single-cell time courses
cell 1
cell 2
cell 3
data
model fit
Cell-specific signaling protein concentrations

**Figure 3**

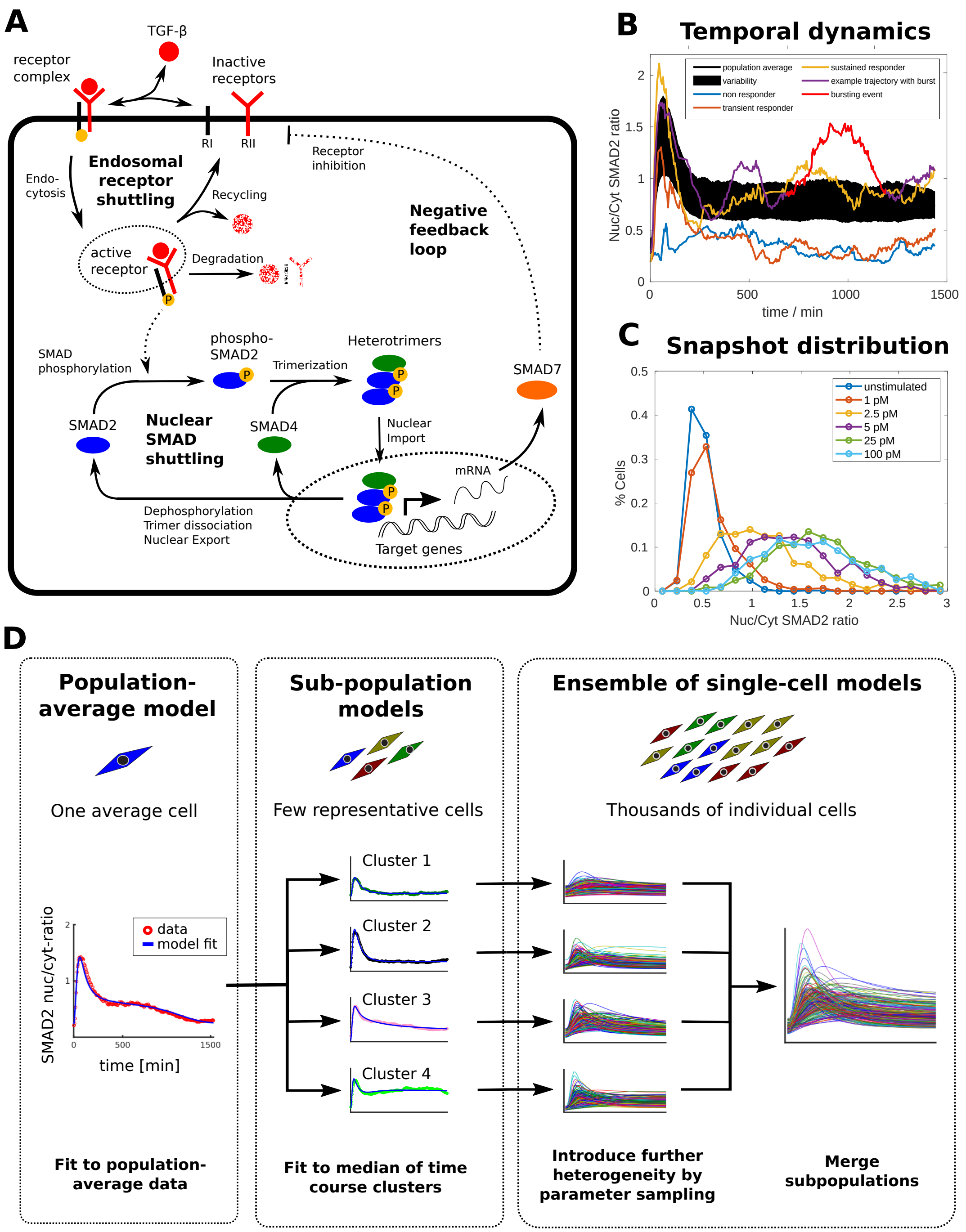